\pdfoutput=1

\documentclass[final,5p,times,twocolumn]{elsarticle}

\usepackage{amssymb}

\biboptions{sort&compress}

\journal{arXiv}

\begin{document}

\begin{frontmatter}

\title{Effect of large-angle incidence on particle identification performance for light-charged ($Z \le 2$) particles by pulse shape analysis with a pad-type nTD silicon detector}

\author[soriko]{Shoichiro Kawase\corref{cor1}}
\ead{kawase@aees.kyushu-u.ac.jp}
\cortext[cor1]{Corresponding author.}

\author[soriko]{Takuya~Murota}
\author[soriko]{Hiroya~Fukuda}
\author[soriko]{Masaya~Oishi}
\author[soriko]{Teppei~Kawata}
\author[soriko]{Kentaro~Kitafuji}
\author[soriko]{Seiya~Manabe\fnref{1}}
\fntext[1]{Present address: National Institute of Advanced Industrial Science and Technology (AIST), AIST Tsukuba Central 2, Tsukuba, Ibaraki, 305-8568, Japan}
\author[soriko]{Yukinobu Watanabe}

\affiliation[soriko]{organization={Department of Advanced Energy Science and Engineering, Kyushu University},%
            addressline={6-1 Kasuga-koen}, 
            city={Kasuga},
            state={Fukuoka},
            postcode={816-8580}, 
            country={Japan}}

\author[butsuri]{Hiroki~Nishibata}
\author[butsuri,riken]{Shintaro~Go}
\author[butsuri]{Tamito~Kai}
\author[butsuri]{Yuto~Nagata}
\author[butsuri]{Taiga~Muto}
\author[butsuri]{Yuichi~Ishibashi}
\author[riken,ut]{Megumi~Niikura}
\author[riken]{Daisuke~Suzuki}
\author[riken]{Teiichiro~Matsuzaki}
\author[riken,kek]{Katsuhiko~Ishida}
\author[ut]{Rurie~Mizuno}
\author[cns]{Noritaka~Kitamura}

\affiliation[butsuri]{organization={Department of Physics, Kyushu University},%
            addressline={744 Motooka, Nishi}, 
            city={Fukuoka},
            state={Fukuoka},
            postcode={819-0395}, 
            country={Japan}}

\affiliation[riken]{organization={Nishina Center, RIKEN},%
            addressline={2-1 Hirosawa}, 
            city={Wako},
            state={Saitama},
            postcode={351-0198}, 
            country={Japan}}

\affiliation[ut]{organization={Department of Physics, the University of Tokyo},%
            addressline={7-3-1 Hongo}, 
            city={Bunkyo},
            state={Tokyo},
            postcode={113-0033}, 
            country={Japan}}

\affiliation[kek]{organization={Muon Science Laboratory, High Energy Accelerator Research Organization (KEK)},%
            addressline={2-4 Shirakata}, 
            city={Tokai, Naka},
            state={Ibaraki},
            postcode={319-1195}, 
            country={Japan}}

\affiliation[cns]{organization={Center for Nuclear Study, the University of Tokyo},%
            addressline={2-1 Hirosawa}, 
            city={Wako},
            state={Saitama},
            postcode={351-0198}, 
            country={Japan}}

\begin{abstract}
In recent years, particle discrimination methods based on digital waveform analysis techniques for neutron-transmutation-doped silicon (nTD-Si) detectors have become widely used for the identification of low-energy charged particles.
Although the particle discrimination capability of this method has been well demonstrated for small incident angles, the particle discrimination performance may be affected by changes in the detector response when the detector is moved closer to the charged particle source and the incident position distribution and incident angle distribution to the detector become wide.
In this study, we performed a beam test for particle discrimination in light-charged ($Z \le 2$) particles using the digital waveform analysis method with a pad-type nTD-Si detector and investigated the dependence of the performance of the particle discrimination on the incident position and incident angle.
As the incident angle increased, a decrease in the maximum current was observed, which was sufficient to affect the performance of the particle discrimination.
This decrease can be expressed as a function of the penetration depth of the charged particles into the detector, which varies for each nuclide.
\end{abstract}


\end{frontmatter}

\section{Introduction}\label{sec:introduction}

Particle identification (PID) is an essential process in nuclear reaction measurements.
In addition to the conventional energy loss ($\Delta E$) and total energy deposit ($E$) method, PID methods based on the digital pulse-shape analysis (DPSA) of neutron-transmutation-doped silicon (nTD-Si) detectors have been developed in recent years~\cite{duenas2012,duenas2013,mengoni2014,assie2015,mahata2018,assie2018,dormard2021} and are becoming more common due to increasing sophistication of data acquisition systems.

The DPSA method has the following advantages over the $\Delta E$-$E$ method.
There is no need to use two detectors as required by the $\Delta E$-$E$ method, and only a single detector is sufficient for PID.
In particular, when identifying low-energy particles, the $\Delta E$-$E$ method requires a rather thin, fragile detector that must be handled with great care. In contrast, the DPSA method requires only one thick detector.
Instead, a waveform digitizer capable of acquiring waveform information is essential for the DPSA method.

Highly segmented detectors such as double-sided silicon strip detectors (DSSSD) are widely used in modern nuclear physics experiments, in which a high angular resolution of the scattered particles is usually required. However, the increased number of channels makes such a detector and the signal readout electronics more costly. On the other hand, pad-type silicon detectors have the advantage of a single detector being able to cover a large solid angle at the cost of not obtaining position and angle information. The small number of channels requires simple and inexpensive data acquisition electronics. This advantage is still of great benefit for certain types of measurements that do not require precise angular information --- for example, measurements of charged particle emission after the muon nuclear capture reaction~\cite{measday2001}.

In such experiments, one should place the detector closer to the target and cover a large solid angle in order to obtain more statistics.
Consequently, charged particles arriving at a large incident angle will also be detected.
Under such conditions, the dependence of the detector response on the position and incident angle can be critical for PID compared to the case of small-angle incidence on a highly segmented detector.
In this study, we experimentally investigated the effect of these factors on the particle discrimination performance of a pad-type nTD-Si detector for light-charged ($Z \le 2$) particles.

\section{Experiment}\label{sec:experiment}

The experiment was conducted using an 8-MV tandem accelerator at the Center for Accelerator and Beam Science (CABAS) of Kyushu University.
To generate light-charged particles with a proton number of two or less, a 24-MeV $^{7}$Li beam irradiated a thin aluminum target with a thickness of 11~$\mu$m.
\begin{figure}[t]
 \centering
 \includegraphics[keepaspectratio, width=\linewidth]
      {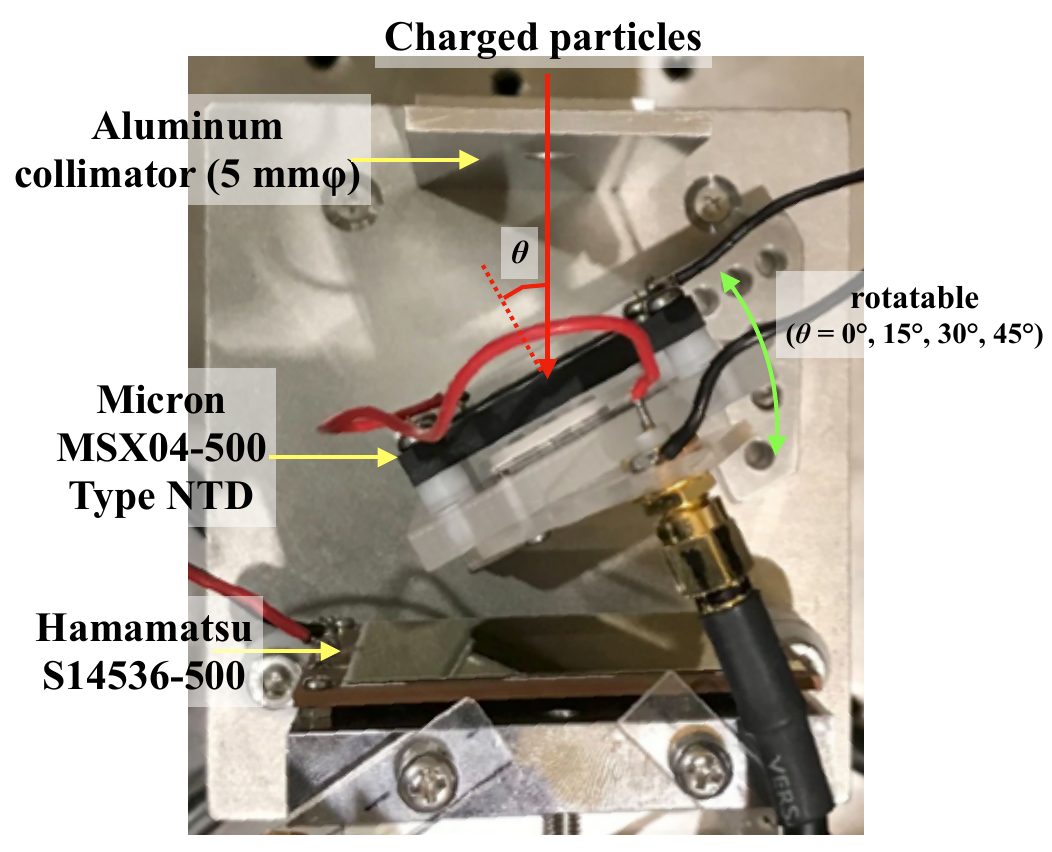}
 \caption{Overview of the detector telescope used in the experiment.}\label{fig:setup}
\end{figure}
Light ions produced by nuclear reactions at the target were detected by using a detector telescope placed at 30$^\circ$ relative to the beam direction.
Both the reaction target and the telescope were placed in a vacuum chamber.
Figure~\ref{fig:setup} presents an overview of the detector telescope. A 496-$\mu$m thick nTD-Si pad detector (Micron Semiconductor MSX04-500 Type NTD) is mounted on the front of the telescope, with the ohmic side facing the target, to take advantage of the particle discrimination performance. It has a sensitive area of $20 \times 20$~mm$^2$.
A 500-$\mu$m-thick veto Si detector (Hamamatsu Photonics S14536-500) with a sensitive area of $48 \times 48$~mm$^2$ is placed on the back of the telescope to detect punch-through particles.
A 3-mm-thick aluminum collimator with a 5-mm$\phi$ hole was placed in front of the nTD-Si to limit the injection area in order to investigate the position dependence of the detector response.
The nTD-Si detector can be rotated around the vertical axis by 0$^\circ$, 15$^\circ$, 30$^\circ$, and 45$^\circ$ with the center of the sensitive area fixed, thereby enabling the incident angle of the light ions onto the detector to be varied.
The detector output signals were amplified by using a 16-channel charge-sensitive preamplifier (Fuji-diamond international co., ltd.~0380-16), with a charge gain of 0.5~V/pC, installed outside the vacuum chamber. It is the register feedback type and its time constant is 51~$\mu$s.
The preamplified signals were then digitized by using a 14-bit waveform digitizer (CAEN SpA V1730SB) with a sampling frequency of 500~MHz. The signals from nTD-Si and the veto Si were acquired independently by the self-trigger with a threshold corresponding to a 7-mV signal input. The recording length and the pregate length with respect to the trigger timing were set to 10~$\mu$s (5000 samples) and 2~$\mu$s (1000 samples), respectively.

\begin{figure}[t]
 \centering
 \includegraphics[keepaspectratio, width=\linewidth]
      {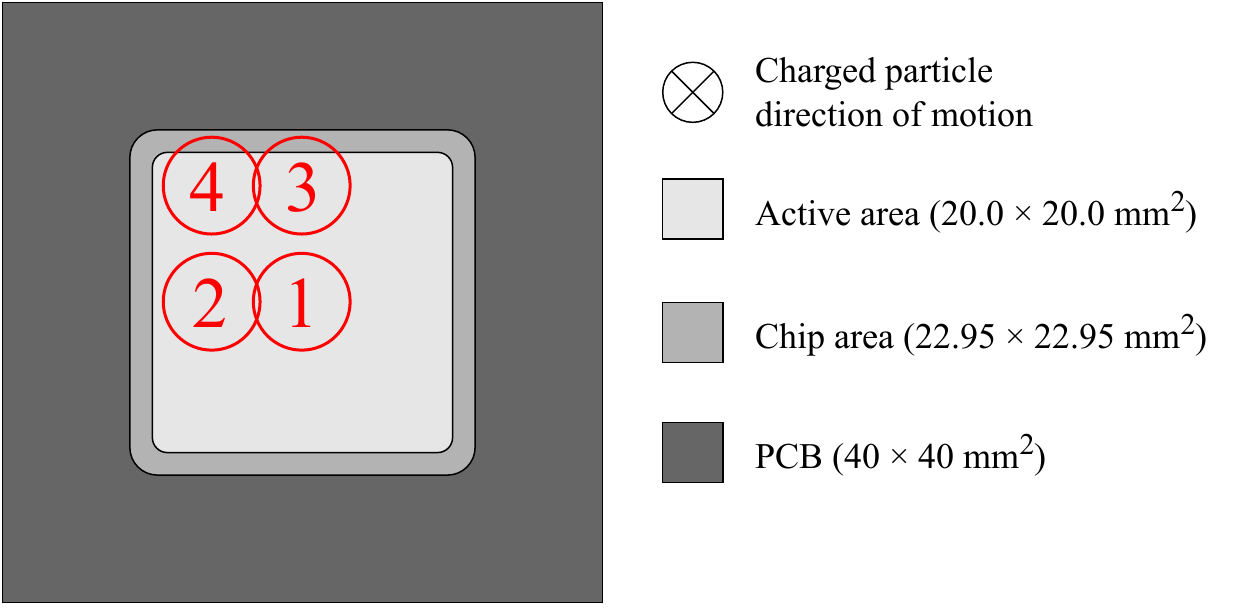}
 \caption{Schematic drawing of the charged-particle injection areas in the nTD-Si detector. Each of the four circles represents an incident area: (1) center, (2) left edge, (3) top edge, and (4) corner.}\label{fig:incident_position}
\end{figure}

In the present experiment, measurements of voltage dependence, position dependence, and incident angle dependence were performed.
First, measurements were made with different bias voltage settings to determine the optimal bias voltage.
Next, the detector response was measured for thecharged particles incident on the four areas --- (1) center, (2) left edge, (3) top edge, and (4) corner, as shown in Fig.~\ref{fig:incident_position} using an aluminum collimator.
In order to investigate the effect on the active area boundary, the top edge and corner areas extend outward from the active area by approximately 1 mm.
Finally, measurements were made at four different angles (0$^\circ$, 15$^\circ$, 30$^\circ$, and 45$^\circ$) of the nTD-Si detector to investigate the dependence of the detector response on the incident angle of the charged particles.
The incident position was fixed at the center throughout the angle dependence measurement.

\section{Data analysis and discussions}\label{sec:analysis}
\begin{figure}[t]
 \centering
 \includegraphics[keepaspectratio, width=0.6\linewidth]
      {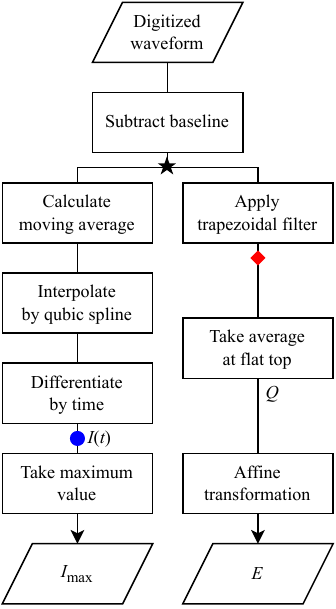}
 \caption{Flow of DPSA. The symbols in the diagram correspond to the waveforms of the same color, as depicted in Fig.~\ref{fig:psa_example}.}\label{fig:psa_flowchart}
\end{figure}
\begin{figure}[t]
 \centering
 \includegraphics[keepaspectratio, width=\linewidth]
      {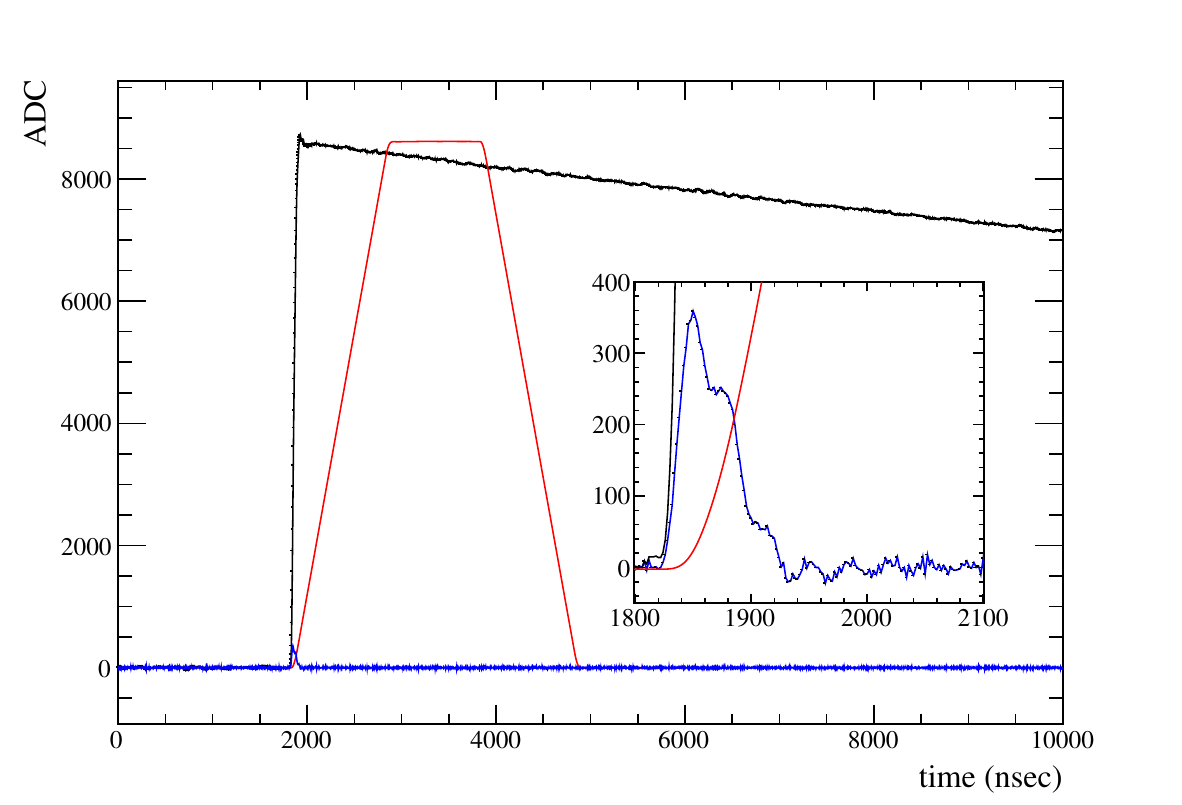}
      \caption{Example of waveforms during DPSA. The black, blue, and red lines represent the digitized waveform after the baseline subtraction, the output of the trapezoid filter, and the time derivative of the moving-averaged waveform, respectively.
        A magnified view close to the rising edge of the signal is displayed in the inner panel.}\label{fig:psa_example}
\end{figure}
\subsection{Pulse shape analysis}

Particle identification by DPSA of the preamplifier output was performed similarly as in Ref.~\cite{mengoni2014}.
The flow of DPSA and example waveforms during the analysis are in depicted Figs.~\ref{fig:psa_flowchart} and \ref{fig:psa_example}, respectively.
The symbols in Fig.~\ref{fig:psa_flowchart} correspond to the waveforms of the same color, as displayed in Fig.~\ref{fig:psa_example}. In the first step, the baseline of the digitized signal of Si detectors, calculated by averaging the first 800 samples, was subtracted from the waveform.
An example of a baseline-subtracted waveform is indicated by the black line in Fig.~\ref{fig:psa_example}.
In order to identify light ions, two feature values were extracted from the digitized waveforms.
One is the maximum current ($I_\mathrm{max}$) and the other is the energy deposit ($E$), as explained in the following subsections.

\subsubsection{Maximum current}

In this experiment, a charge-sensitive preamplifier was used to amplify the signal from the silicon detector. Therefore, the following procedure was used to derive $I_\mathrm{max}$ for each waveform.
First, a moving average of the five data points, which functions like a low-pass filter, was taken to reduce the effect of background fluctuations.
Next, the discrete data points with a 2-ns sampling period were interpolated with a cubic spline to make it continuous and then differentiated over time. Since the preamplified waveforms have charge information, their time derivative correspond to the current information.
The example of the time-differentiated waveform is indicated by the blue line in Fig.~\ref{fig:psa_example}.
The maximum value of the differentiated waveform was taken as $I_\mathrm{max}$.

\subsubsection{Energy deposit}

The energy deposit ($E$) of the incident particle in the detector was obtained from the total charge $Q$ of a detector pulse.
For the compensation of the exponential decay of the preamplified signal during the rise time, a trapezoidal filter~\cite{trapezoid1, trapezoid2} was applied to the pulse.
The rise time, the flat top, and the pole zelo of the trapezoidal filter were set to 1~$\mu$s, 1~$\mu$s, and 44~$\mu$s, respectively.
The waveform after application of the trapezoidal filter is indicated by the red line in Fig.~\ref{fig:psa_example}.
The charge $Q$ was obtained by averaging 16 samples in the flat top of the trapezoid.
The energy deposit $E$ was then calculated by a linear transformation of $Q$.
The energy calibration was performed using the punch-through energy of protons, deuterons, and tritons at 0$^\circ$ and a mixed alpha source consisting of three alpha emitters --- $^{148}$Gd, $^{241}$Am, and $^{244}$Cm. The energy resolution was a maximum of 32 keV (FWHM) for these three alpha peaks.

\subsubsection{Pileup rejection}
Suppose a signal arrives when the exponentially decaying tail of the previous signal is still present. In such a case, DPSA cannot be appropreately processed because the baseline calculation is incorrect.
The maximum event rate for nTD-Si in this measurement was approximately 100~cps. Although the rate of pileup events was not high, they were removed to achieve the best particle discrimination performance.
In the current analysis, events whose time stamp difference from the preceding events was less than 1~ms, which is much longer than the preamplifier's time constant of 51~$\mu$sec, were excluded as pileup events.

\subsubsection{Particle identifiaction}

\begin{figure}[t]
 \centering
 \includegraphics[keepaspectratio, width=\linewidth]
      {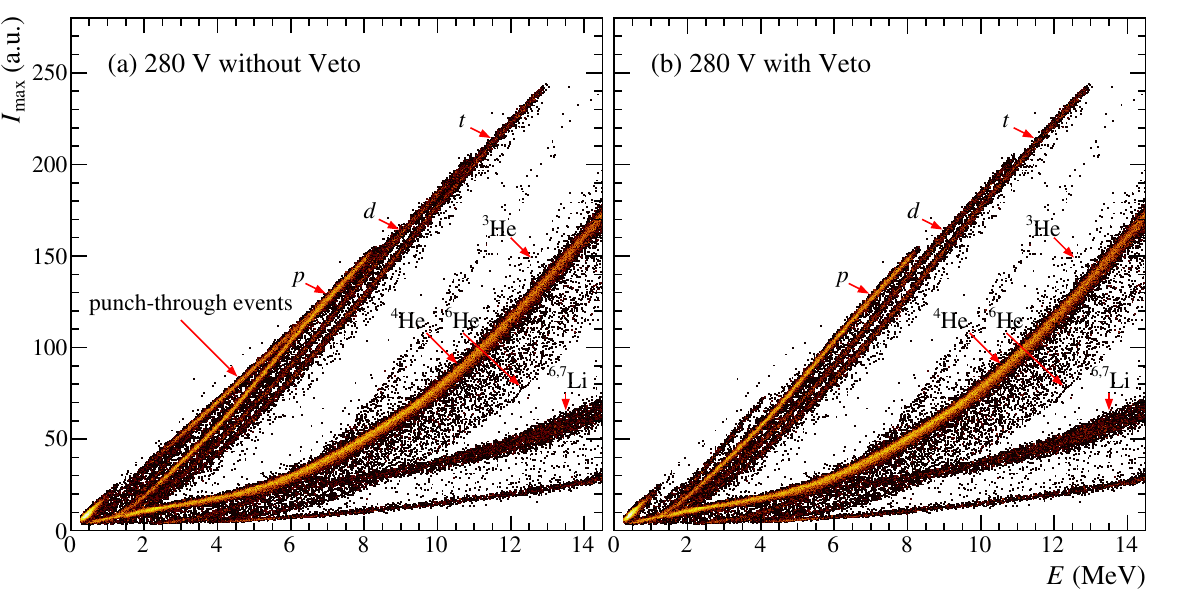}
 \caption{Particle identification plots with a bias voltage of 280~V: (a) no rejection of punch-through events and (b) with rejection of punch-through events. }\label{fig:pid}
\end{figure}
In the DPSA method, PID is performed through the correlation between $E$ and $I_\mathrm{max}$.
Figure~\ref{fig:pid}(a) depicts the correlation between $E$ and $I_\mathrm{max}$, henceforth referred to as a PID plot, with the bias voltage of 280~V.
A figure similar to that in the literature was obtained, and the loci corresponding to protons, deuterons, tritons, and alpha particles are clearly recognized.
Other events corresponding to $^{3}\mathrm{He}$, $^{6}\mathrm{He}$, and $^{6,7}\mathrm{Li}$ are also evident, although the separation is not clear at this stage.
The line that folds back to the lower left from the high-energy end of the proton locus is due to events in which particles had penetrated the nTD-Si.

\subsubsection{Elimination of punch-through events}

In order to eliminate events where a charged particle punched through the nTD-Si,
events with a time difference of 100~ns between the time stamp of the nTD-Si signal and that of the veto Si signal were excluded.
The PID plot after a rejection of punch-through events is displayed in Fig.~\ref{fig:pid} (b).
The penetrating particle events extending to the lower left of the point of maximum energy deposit are correctly excluded after the rejection. The separation in the high-energy region is improved, and the hydrogen isotopes are clearly distinguished.
Hereafter, all the punch-through events are eliminated in the analysis.

\subsection{Selection of optimum bias}

It has been found that particle performance improves more at lower biases as compared to the over-depletion bias of the nTD-Si detector~\cite{duenas2013}.
In this study, measurements were also made at various bias voltage settings to investigate the optimal voltage for light ion discrimination.

\begin{figure}[t]
 \centering
 \includegraphics[keepaspectratio, width=\linewidth]
      {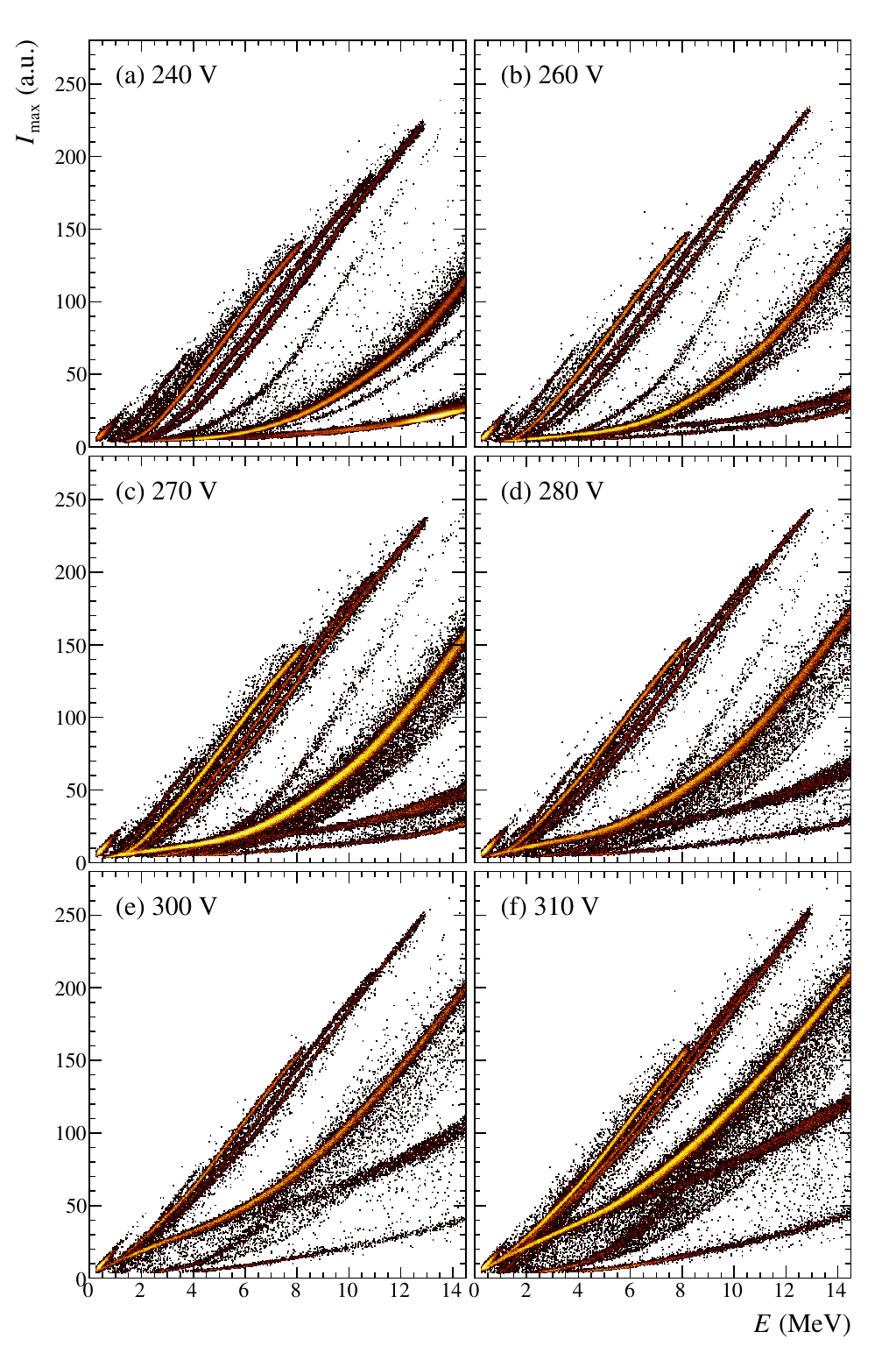}
 \caption{Particle identification plots for light-charged particles with various bias voltages.}\label{fig:pid_murota}
\end{figure}

Figure~\ref{fig:pid_murota} depicts the PID plots for different bias voltages.
As the bias voltage increases, $I_\mathrm{max}$ tends to increase for the same energy loss. This is because a higher bias voltage shortens the charge collection time.
As the bias is lowered from 300~V, which is the nominal full-depletion bias of the MSX04-500, the PID performance improves, particularly in the low-energy region.
On the other hand, the measurement at 310~V bias shows poor particle discrimination for low-energy hydrogen isotopes.

In order to quantitatively evaluate the particle discrimination performance under each measurement condition, the figure of merit (FOM) was defined in the same manner as done in Ref.~\cite{assie2018}:
\begin{equation}
  \mathrm{FOM}_{pd} = \frac{2\left|\mu_p - \mu_d\right|}{2.35(\sigma_p + \sigma_d)},
\end{equation}
where $\mu$ and $\sigma$ are the mean and standard deviation of the distribution of $I_\mathrm{max}$ for a given energy loss by a given particle, respectively.
Using this definition, an FOM greater than 1.5 can generally be considered to have sufficient discrimination power.

\begin{figure}[t]
 \centering
 \includegraphics[keepaspectratio, width=\linewidth]
      {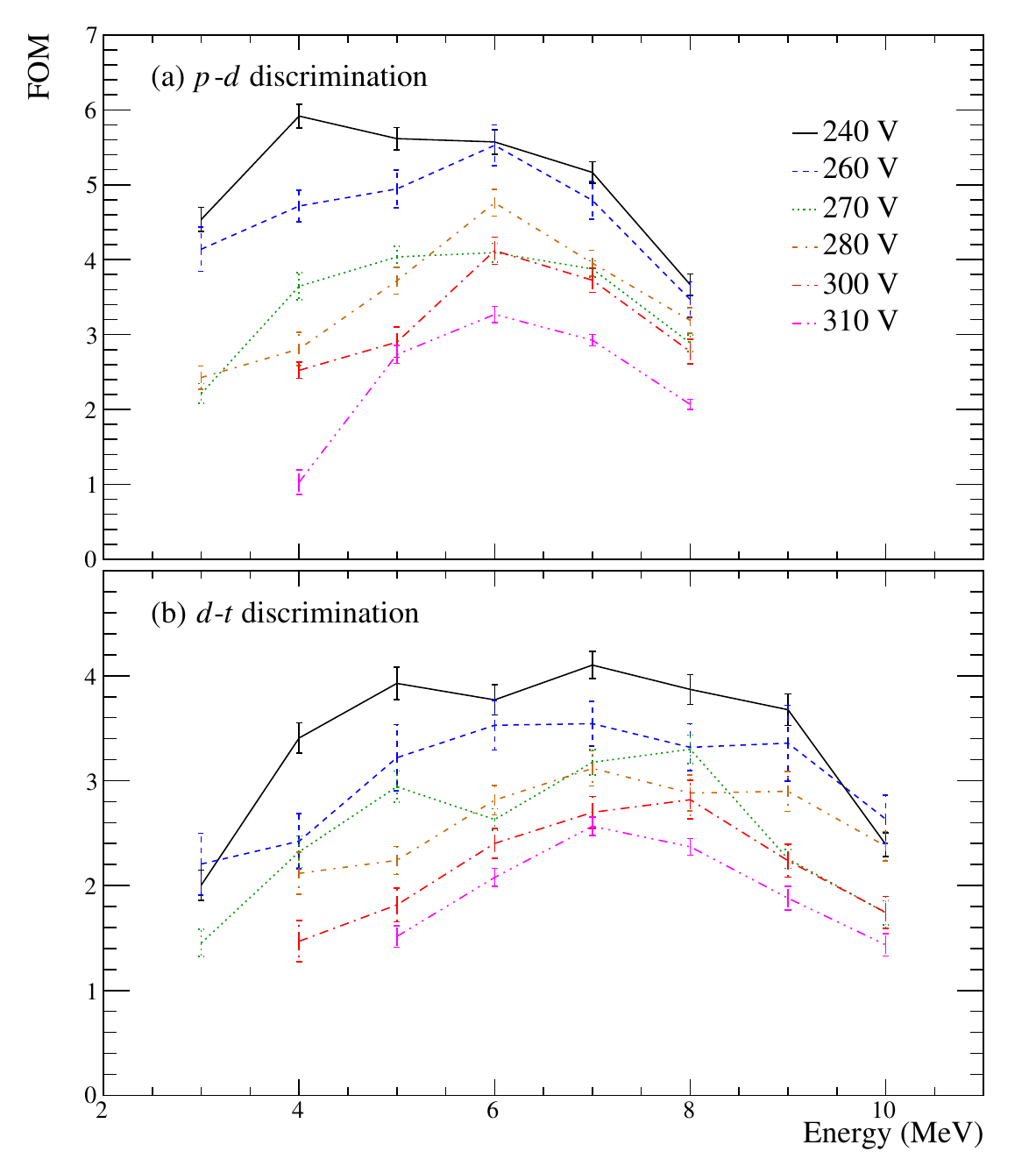}
 \caption{Figure of merits for (a) $p$-$d$ discrimination and (b) $d$-$t$ discrimination with various bias voltages.}
 \label{fig:fom}
\end{figure}

Figures~\ref{fig:fom} (a) and (b) depict FOMs for $p$-$d$ and $d$-$t$ discrimination for various bias voltages and energy losses.
Points corresponding to conditions in which the two particles were indistinguishable are not plotted.
As a general trend, smaller bias voltages improved the performance of particle discrimination for hydrogen isotopes, and satisfactory discrimination was achieved for hydrogen ions of 3 MeV or higher at 260~V or lower. However, if the bias voltage is set too low, the charge collection efficiency will also decrease and the signal wave height will also be low, which may cause low-energy particles to be missed. In fact, a decrease in the pulse height and deterioration in energy resolution was observed at 240~V for the alpha particles used in the energy calibration. Therefore, 260~V was selected as the optimal bias in this study, and all subsequent measurements hereafter were made at 260~V bias.

\subsection{Position dependence}

\begin{figure}[t]
 \centering
 \includegraphics[keepaspectratio, width=\linewidth]
      {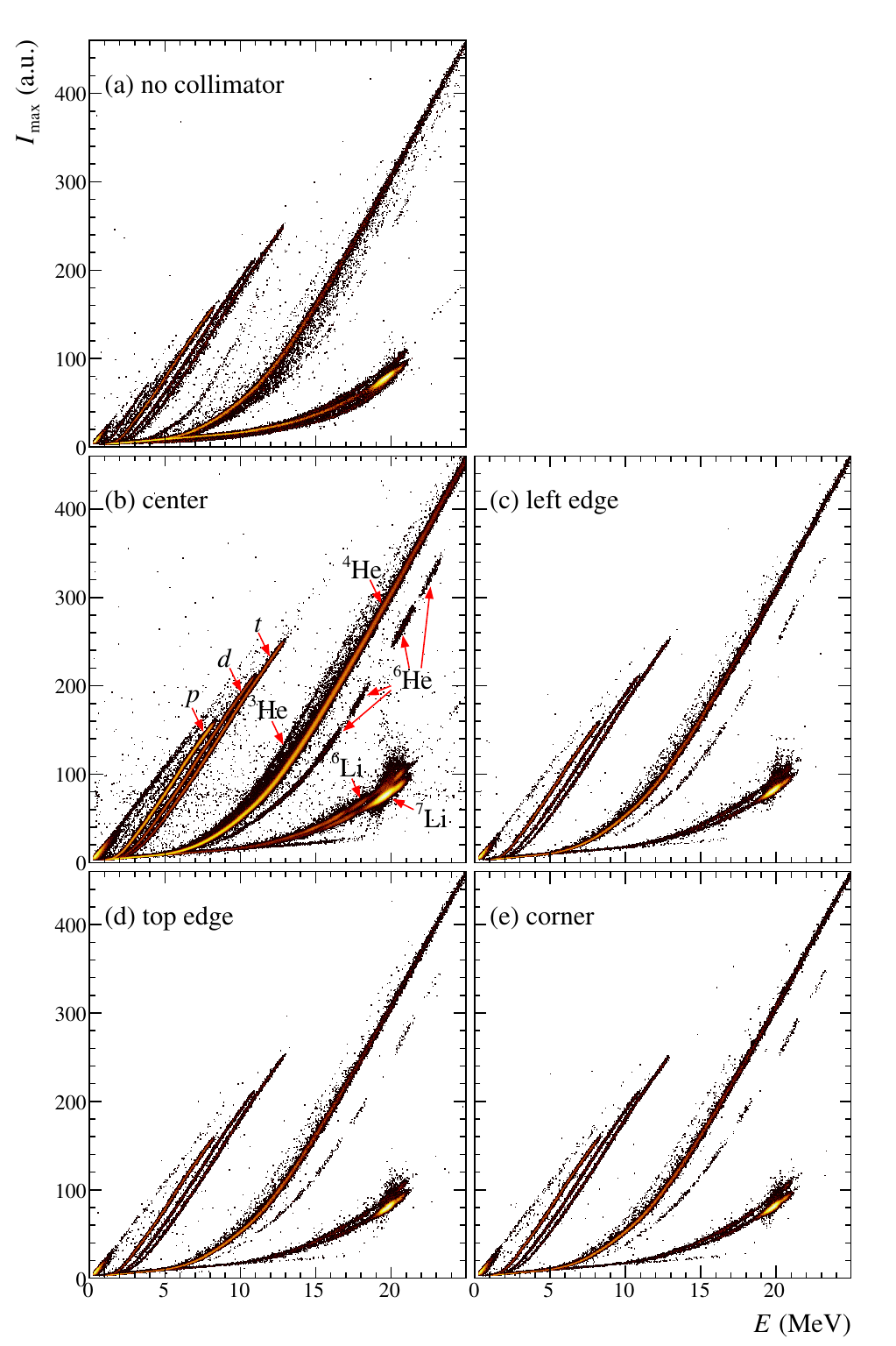}
 \caption{Particle identification plots for light-charged particles with various incident regions.}
 \label{fig:pid_posdep}
\end{figure}

The dependence of the particle discrimination performance on the incident position is discussed on the basis of the results of measurements when charged particles are injected into the four regions depicted in Fig.~\ref{fig:incident_position} using the collimator and when charged particles are irradiated without the collimator --- that is, the entire detector is irradiated with charged particles.
Figure~\ref{fig:pid_posdep} presents the PID plots for different incident regions.
For the measurements with the collimator, there is little difference in the PID plot, except for the difference in statistics.
The locus between the hydrogen isotope loci and the helium isotope loci present when the collimator was absent is eliminated by using the collimator to limit the incident area.
In addition, the use of the collimator significantly reduces the smearing of events from the respective loci to the low $I_\mathrm{max}$ side.
This limitation of the incident area improves the discrimination performance, particularly between $^{4}$He and $^{6}$He in the energy region below 15~MeV.
The absence of such events, even when charged particles are incident not only in the central area but also in the edge and the corner areas, suggests that these events occur when particles are incident at the very edge of the chip area of the detector.

\begin{figure}[t]
 \centering
 \includegraphics[keepaspectratio, width=\linewidth]
      {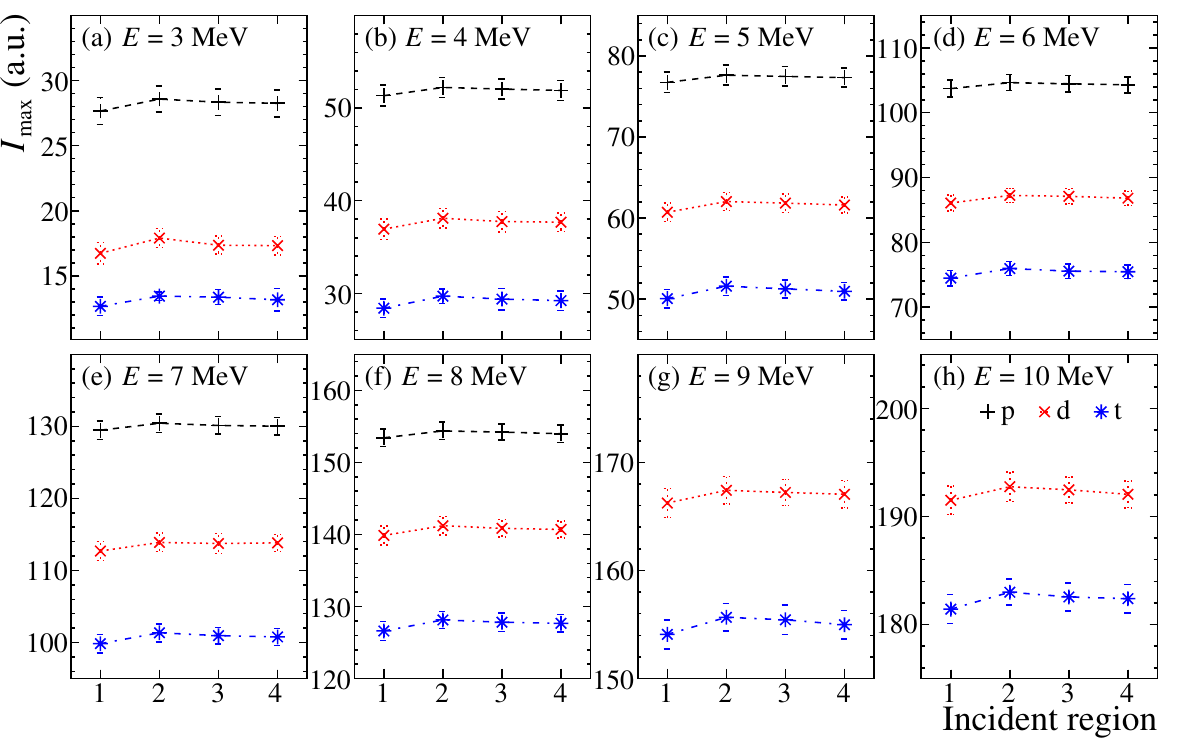}
 \caption{Variation of $I_\mathrm{max}$ for hydrogen isotopes with various incident regions. The error bars represent standard errors of the $I_\mathrm{max}$ distributions.}
 \label{fig:i_posdep}
\end{figure}

Figure~\ref{fig:i_posdep} depicts the variation of $I_\mathrm{max}$ for different incident areas for different incident energies. Each symbol and error bar represents the mean and standard error of the $I_\mathrm{max}$ distribution. 
For all energy regions, $I_\mathrm{max}$ is slightly smaller when particles are incident at the center. The reason for this is not clear, but it will not be discussed in this paper because this decrese does not affect the particle identification performance that is of interest here.

\subsection{Incident angle dependence}

\begin{figure}[t]
 \centering
 \includegraphics[keepaspectratio, width=\linewidth]
      {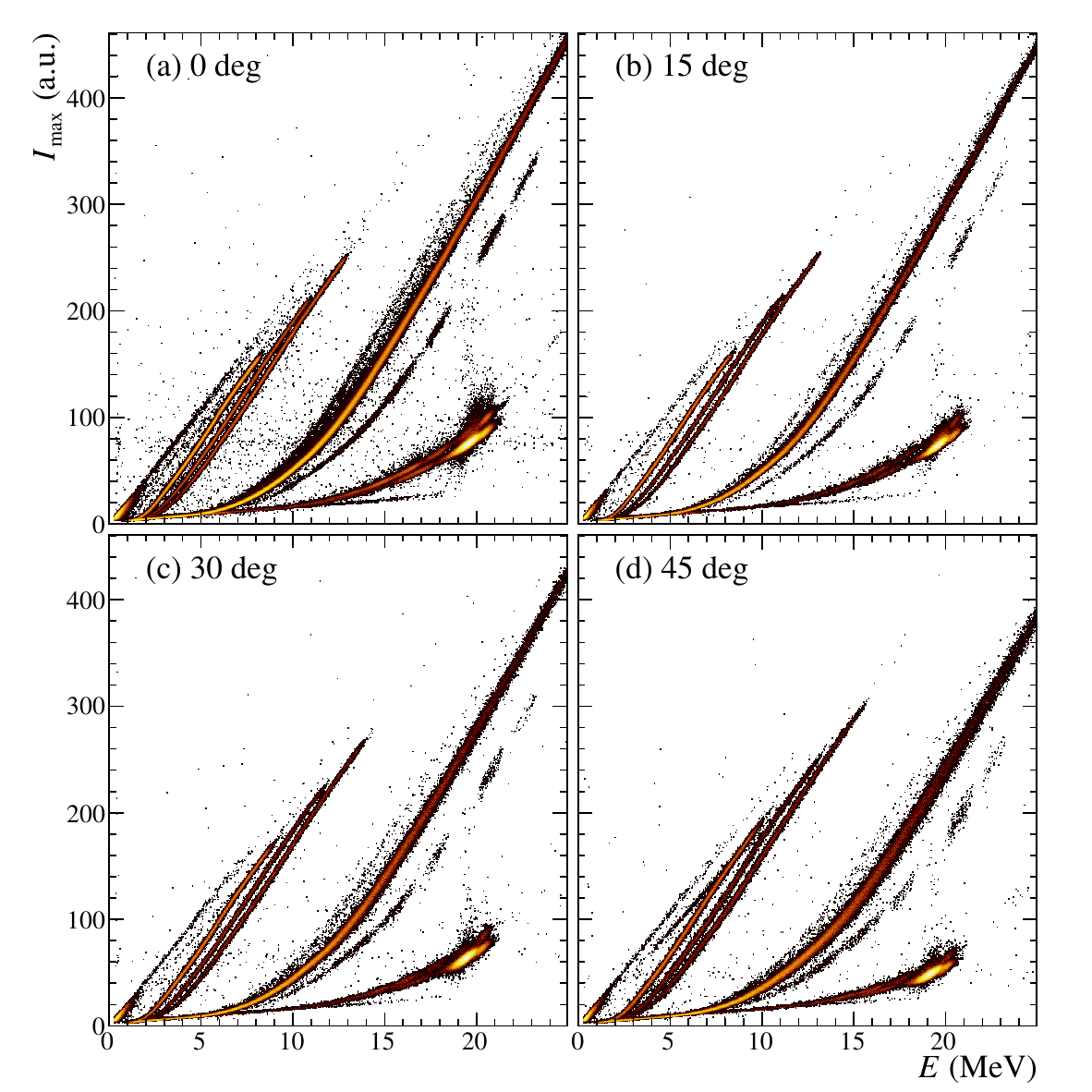}
 \caption{Particle identification plots for light-charged particles with various incident angles.}
 \label{fig:pid_angdep}
\end{figure}

\begin{figure}[t]
 \centering
 \includegraphics[keepaspectratio, width=\linewidth]
      {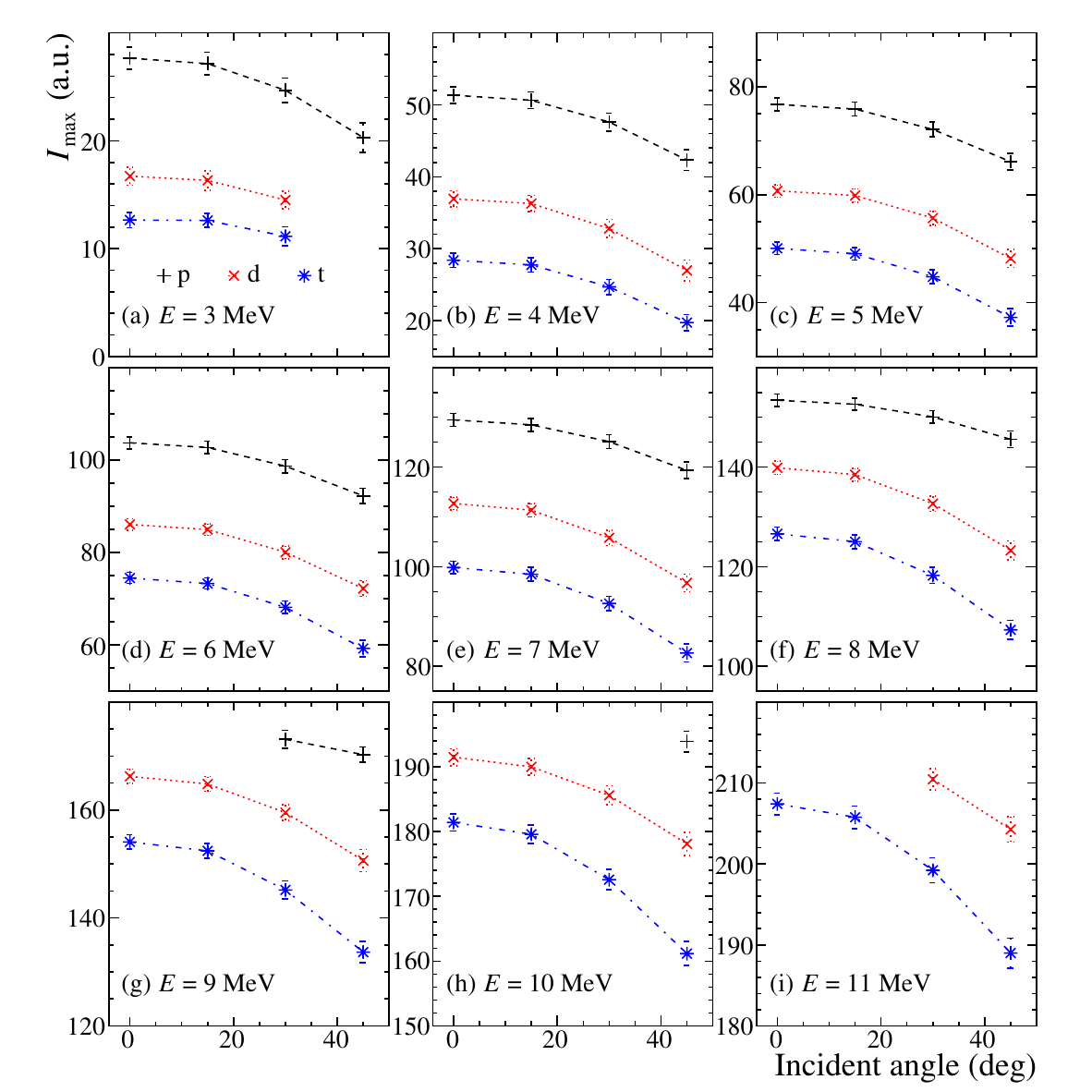}
 \caption{Variation of $I_\mathrm{max}$ for hydrogen isotopes with various incident angles.  The error bars represent standard errors of the $I_\mathrm{max}$ distributions.}
 \label{fig:i_angdep}
\end{figure}

Figure~\ref{fig:pid_angdep} depicts the PID plots for incident angles of 0$^\circ$, 15$^\circ$, 30$^\circ$, and 45$^\circ$.
For the incident angle at which the measurements were made, it was found that particle discrimination was possible even at large angles of 45$^\circ$ if the incident angle distribution was sufficiently restricted.
The maximum energy of stopped particles increases as the incident angle increases because the effective thickness increases by a factor of $1/\cos\theta$.
In general, for particles with the same kinetic energy, $I_\mathrm{max}$ decreases as the incident angle increases.
Figure~\ref{fig:i_angdep} depicts the variation of $I_\mathrm{max}$ for each energy range and for each hydrogen ion as a function of the incident angle.
The higher the neutron number and the lower the energy, the higher the rate of the reduction of $I_\mathrm{max}$.
From these figures, we can conclude that hydrogen ions can be well discriminated by $I_\mathrm{max}$ if the incident angle is limited between 0$^\circ$ and 30$^\circ$. For example, assuming a point source of charged particles, sufficient PID can be achieved if the detector is at least 5 cm away from the source.

\begin{figure}[t]
 \centering
 \includegraphics[keepaspectratio, width=\linewidth]
      {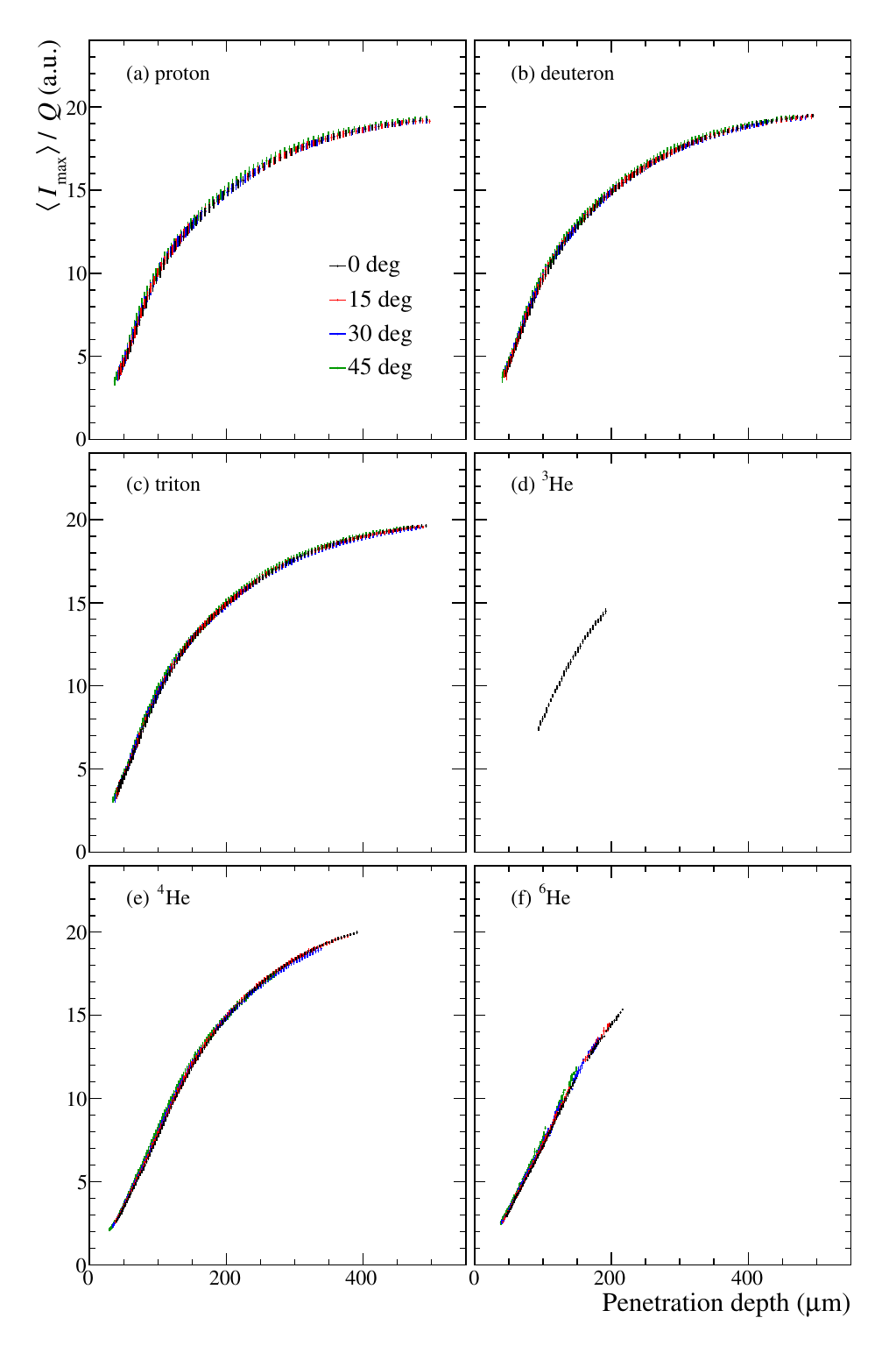}
 \caption{Penetration depth dependence of the $\langle I_\mathrm{max} \rangle$/$Q$ ratio for each nuclide.}
 \label{fig:depth_qi}
\end{figure}

In order to understand how increasing the incident angle reduces $I_\mathrm{max}$, the relationship between the depth of penetration of the charged particle into the detector and the maximum current is discussed.
$I_\mathrm{max}$ is considered to be approximately proportional to the charge collection rate. The charge collection rate is affected by the magnitude of the electric field in the silicon detector and the drift distance to the electrode, both of which are functions of the penetration depth of the detector by the charged particles.
Thus, the ratio $\langle I_\mathrm{max} \rangle$/$Q$ can be considered proportional to the rate of charge collection and independent of the amount of charge.
Figures~\ref{fig:depth_qi} (a)--(f) depict $\langle I_\mathrm{max} \rangle$/$Q$ as a function of penetration depth for each hydrogen and helium isotope.
Here, the penetration depth $d$ was defined in the following manner:
\begin{equation}
d \equiv R(E)\cos\theta,
\end{equation}
where $R(E)$ is the range in silicon as a function of total kinetic energy and was calculated using the methods in Refs.~\cite{hrange} (for H) and \cite{herange} (for He) implemented in the LISE++ code~\cite{lise}.

It is evident that the $\langle I_\mathrm{max} \rangle$/$Q$ ratio for each nuclide is almost independent of the incident angle of charged particles and can be well represented only by the penetration depth.
Once this ratio has been measured for the detector to be used, the detector response can be simulated using the nuclide and its incident angle distribution as input. This enables us to evaluate the particle discrimination performance in future experimental setups.

In addition, this relationship may enable the nTD-Si detector to be used as an incident angle detector. If the incident nucleus is identified, the $\langle I_\mathrm{max} \rangle$/$Q$ ratio can be used to determine the penetration depth and, thus, provide information on the incident angle of low-energy charged particles.
It can be used, for example, to evaluate the background events due to charged particles not originating from the target and to identify the target from which the charged particles are originating from in multiple target measurements.

\begin{figure}[t]
 \centering
 \includegraphics[keepaspectratio, width=\linewidth]
      {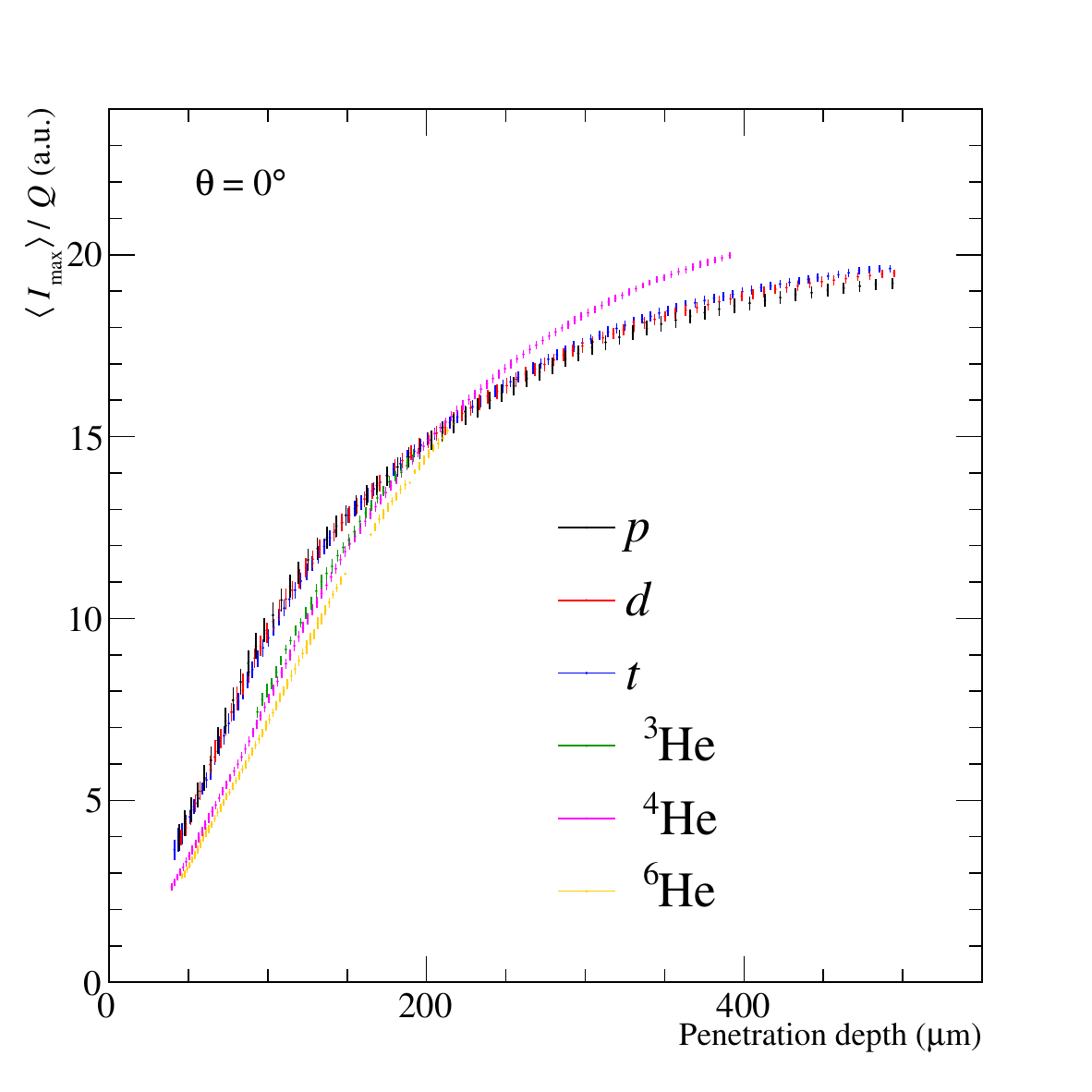}
 \caption{Penetration depth dependence of the $\langle I_\mathrm{max} \rangle$/$Q$ ratio for each nuclide at an incident angle of 0$^\circ$.}
 \label{fig:depth_qi_0deg}
\end{figure}

Figure~\ref{fig:depth_qi_0deg} depicts the $\langle I_\mathrm{max} \rangle$/$Q$ ratio as a function of penetration depth for different nuclides at an incident angle of 0$^\circ$.
The magnitude of the $\langle I_\mathrm{max} \rangle$/$Q$ ratios for hydrogen and helium are distributed in a similar range, but the slope is steeper for helium than that for hydrogen. The slope for nuclides with the same number of protons tends to increase slightly with an increasing mass number, although the difference is smaller than that for different proton numbers.
The dependence on the number of protons and mass number at large penetration depths can be qualitatively discussed because as the number of protons and mass number increases, the width of the Bragg peak narrows and the position distribution of the energy loss becomes closer to the end of the range, thereby resulting in a larger $\langle I_\mathrm{max} \rangle$/$Q$ ratio.
On the other hand, in the shallow penetration depth region, the electric field is weaker due to the under-depletion bias. It is possible that the space charge effect, although small, is present because the charge density generated at the Bragg peak is larger for particles with larger proton and mass numbers, thereby resulting in slightly longer charge collection times.

\section{Summary and conclusions}\label{sec:conclusions}

Particle identification for light-charged particles was performed using a pad-type nTD-Si detector by using the DPSA method, and the incident position and angle dependence of the PID performance was investigated.
No significant position dependence was observed, except at the very edge of the chip area.
On the other hand, with regard to the incident angle dependence, there was a significant decrease in the maximum current $I_\mathrm{max}$ with an increasing incident angle, which was sufficient to affect the PID performance.
It was found that the reduction in $I_\mathrm{max}$ can be well expressed as a function of the penetration depth of the charged particles into the nTD-Si detector.
This result enables the evaluation of PID performance in new measurement systems. It demonstrates the possibility of using an nTD-Si detector as an incident angle detector for low-energy charged particles.

\section*{Declaration of competing interests}

The authors declare that they have no known competing financial interests or personal relationships that could have influenced the work reported in this paper.

\section*{Data availability}

Data will be made available on request.

\section*{Acknowledgments}
This work was supported by JSPS KAKENHI Grant Numbers JP19H05664 and JP21H01863.

\end{document}